\documentclass[aps,prd,nofootinbib,floatfix,superscriptaddress,
preprint
]{revtex4-1}
\usepackage[]{qrcode}
\usepackage{graphicx,subfigure}
\usepackage{amsmath,amssymb,amsfonts}
\usepackage{bm}
\usepackage{color}
\usepackage{lineno}
\usepackage{comment}
\usepackage[hidelinks]{hyperref}

\newcommand{\be}{\begin{equation}}
\newcommand{\ee}{\end{equation}}

\begin{document}

\preprint{MIT-CTP/4888}
\preprint{PUPT-2518}
\vspace*{3cm}

\title{Anomalous Transport and Generalized Axial Charge}

\author{Vladimir P. Kirilin}
\email{vkirilin@princeton.edu}
\affiliation{Department of Physics, Princeton University, Princeton, NJ 08544}
\affiliation{ITEP, B. Cheremushkinskaya 25, Moscow, 117218, Russia}

\author{Andrey V. Sadofyev}
\email{sadofyev@mit.edu}
\affiliation{Center for Theoretical Physics, Massachusetts Institute of Technology, Cambridge, MA 02139, USA}
\affiliation{ITEP, B. Cheremushkinskaya 25, Moscow, 117218, Russia}

\begin{abstract}
In this paper we continue studying the modification of the axial charge in chiral media by macroscopic helicities. Recently it was shown that magnetic reconnections result in a persistent current of zero mode along flux tubes. Here we argue that in general a change in the helical part of the generalized axial charge results in the same phenomenon. Thus one may say that there is a novel realization of chiral effects requiring no initial chiral asymmetry. The transfer of flow helicity to zero modes is analyzed in a toy model based on a vortex reconnection in a chiral superfluid. Then, we discuss the balance between the two competing processes effect of reconnections and the chiral instability on the example of magnetic helicity. We argue that in the general case there is a possibility for the distribution of the axial charge between the magnetic and fermionic forms at the end of the instability.
\end{abstract}

\date{\today}
\maketitle

\section{Introduction}

Physics of chiral media has attracted a lot of attention recently. Such systems have massless fermions as basic constituents and thus the underlying microscopic theory is anomalous. The anomaly manifests itself as a breaking of the classical symmetry, corresponding to the axial charge conservation, via loop contributions in the presence of external fields \cite{Adler:1969gk, Bell:1969ts}. It is by now well established that the anomaly, a purely quantum phenomenon, has macroscopic manifestations for the transport in the media (for review see \cite{Kharzeev:2012ph, Zakharov:2012vv}). More specifically, one observes anomalous transport of electric and axial charges:
\begin{eqnarray}
\label{CE}
&~&\vec{j}_{V}=\frac{1}{2\pi^2}\mu_{A}\vec B+\frac{1}{\pi^2}\mu_V\mu_A\vec\Omega\notag\\
&~&\vec{j}_A=\frac{1}{2\pi^2}\mu_{V}\vec B+\left(\frac{\mu_V^2+\mu_A^2}{2\pi^2}+\frac{T^2}{6}\right)\vec\Omega\,,
\end{eqnarray}
where $\mu_{V(A)}=\frac{1}{2}(\mu_R\pm\mu_L)$ is vector(axial) chemical potential (R/L refers to right/left-handed fermions), $B$ is the magnetic field and $\Omega$ is the angular velocity of the matter. These phenomena are widely discussed in the literature \cite{Kharzeev:2015znc} and could take place in various systems -- from quark-gluon plasma at high temperature to primordial plasma and Dirac/Weyl semimetals.

It is well known that in the presence of external fields the axial charge is not conserved due to anomalous divergence of the axial 4-current: $\partial_\mu J_A^\mu=\frac{e^2}{2\pi^2}E\cdot B$. However the right-hand side could be treated as being a divergence of some auxiliary construction $K_\mu=\frac{1}{4\pi^2}\epsilon_{\mu\nu\rho\sigma}A^\nu\partial^\rho A^\sigma$ known as Chern current. Moving this term to the left-hand side one may notice a conserved combination of two contributions. Moreover despite the apparent gauge dependency of $K^\mu$, one may actually obtain a gauge invariant charge:

\begin{eqnarray}
\label{q50}
&~&Q_5=N_5+K^0=N_5+\frac{1}{4\pi^2}\int A\cdot Bd^3 x\,.
\end{eqnarray}
Here $N_5=N_R-N_L$ is the chiral asymmetry, the difference between numbers of right and left particles in the system. From an effective field theory \cite{Sadofyev:2010is} we know that the hydrodynamic expectation value of an operator may be obtained by the chemical shift of the potential $A_\nu\to A_\nu+\mu u_\nu$  of its vacuum form which is originating in a naive relativization $\mu\bar\psi\gamma^0\psi\to \mu u_\mu\bar\psi\gamma^\mu\psi$ where $u_\mu$ is treated as a constant boost and we ignore its gradients for now. Applying this procedure to (\ref{q50}) we find the generalized axial charge including flow counterparts \cite{Avdoshkin:2014gpa}

\begin{eqnarray}
\label{q5}
Q_5=N_5+\frac{1}{4\pi^2}\int A\cdot Bd^3x+\frac{1}{2\pi^2}\int \mu^2v\cdot\Omega d^3x+\frac{1}{2\pi^2}\int \mu v\cdot Bd^3x\,,
\end{eqnarray}
where macroscopic contributions are referred as helicities\footnote{We will often use notations $\mathcal{H}_{mh}=\int A\cdot Bd^3x$, $\mathcal{H}_{fh}=2\int \mu^2v\cdot\Omega d^3x$ and $\mathcal{H}_{mfh}=2\int \mu v\cdot Bd^3x$ for magnetic, flow and mixed (magnetic-flow) helicities respectively}. It is worth mentioning that all three macroscopic contributions are of topological nature and represent linking number between field/flow lines \cite{Moffatt}. This unification of topological quantities is a pleasurable conjecture which still should be formalized by a microscopic consideration, see e.g. \cite{Avkhadiev:2017fxj}.

Recently it was argued that a change in magnetic helicity, due to reconnections, produces zero modes and generates a persistent current along flux tube \cite{Hirono:2016jps}. This current is  an analogue of chiral magnetic effect (CME) by its chiral nature and orientation along magnetic field. However it exists even in the absence of an initial asymmetry or, equivalently, with $\mu_5=0$. Here we stress that this process may be seen as the axial charge conservation (\ref{q50}) and extended to an arbitrary source of magnetic helicity change. Thus this phenomenon provides a direct check of the axial charge modification. Indeed, in the absence of the anomaly two contributions are completely separate while the generation of zero modes by a change in helicity is a feature of the modified conservation law.

We employ this idea to probe the suggested unification of the axial charge by vortical helicities. We start with the possible vortical effect which corresponds to the zero mode current generated along hydrodynamic vortices in the full analogy with \cite{Hirono:2016jps}. Consideration of chiral vortical effect (CVE) is complicated by the absence of a microscopic description for the fluid interaction with fermionic modes. Moreover fermions are of purely quantum nature and it is impossible to keep them in the hydrodynamic limit. To avoid these issues we concentrate on the case of a superfluid at zero temperature where the macroscopic velocity field has a clear microscopic counterpart - the Goldstone boson. While superfluid flow is irrotational it is known that a state with nonzero angular momentum may appear to be energetically preferable \cite{landau1959fluid} and the system tends to generate defects: superfluid vortices. A field theoretical description of a relativistic superfluid and its defects is a textbook topic (for recent discussion see, e.g., \cite{Son:2002zn, Nicolis:2011cs}). In the case of a chiral superfluid CVE could be generated only along vortices \cite{Kirilin:2012mw}. We study interaction of zero modes with the defects through a helicity changing process, say reconnection of superfluid vortices, and show the generation of the current. It is then argued that this result may be extended to the case of a general change in the full helicity leading to a wider set phenomena. Thus in a chiral superfluid one can assign some axial charge to a linked configuration of vortices and/or magnetic field lines. This is the main result of our consideration here.

 We then discuss the macroscopic current caused by magnetic reconnections in a medium with finite conductivity. The resulting current is proportional to the total change of the helicity from the initial state and directed along magnetic field. Notice that the usual dissipative currents are also induced by the reconnection process, whereas the zero mode component of the current is nondissipative, see, e.g., \cite{Hirono:2016jps}, and thus naively only limited by the magnitude of the initial helicity. We stress however that the zero mode production by magnetic reconnections has a competitor -- chiral magnetic instability \cite{Boyarsky:2012ex, Akamatsu:2013pjd}, generating magnetic helicity out of $\mu_5$. The instability is not a peculiar feature of a system exhibiting CME transport but could be sourced by any current directed along magnetic field (see, e.g., \cite{Moffatt, Yokoi:2013mja}). The final state, the system is brought by the instability, corresponds to some distribution of the axial charge among different forms (see e.g. \cite{Rubakov:1985nk,Joyce:1997uy,Boyarsky:2011uy,Tashiro:2012mf,Akamatsu:2013pjd,Khaidukov:2013sja,Avdoshkin:2014gpa,Hirono:2015rla,Boyarsky:2015faa,Manuel:2015zpa,Tuchin:2016qww,Zakharov:2016lhp}). It is usually supposed that in this state almost all of the axial charge is in the form of the magnetic helicity. In more details, it is shown that the final field configuration corresponds to self-linked Chandrasekhar-Kendall (CK) state of maximal possible size \cite{Hirono:2015rla}. The axial charge is realized through the field helicity while the chiral asymmetry is suppressed. Its value corresponds to the low CME current required to support CK configuration. On the other hand there are basic arguments suggesting a picture of somewhat equal distribution within available degrees of freedom \cite{Zakharov:2016lhp}. We argue here qualitatively that the final distribution depends on a general set of IR parameters (such as the system size, as in \cite{Hirono:2015rla}). In the general case it supports the perspective of \cite{Zakharov:2016lhp} which advocates for the more uniform distribution of the charge between the constituents. Finally, we argue that the picture above can be extended to the case of chiral vortical instability \cite{Avdoshkin:2014gpa}.

\section{Magnetic reconnections}

It is known (see \cite{Nielsen:1983rb}) that the anomalous divergence of the axial current can be understood in terms of zero modes in external magnetic field. We may then classify fermionic excitations in terms of the occupancy numbers of the Landau levels (LLs). There is only one chiral mode which contributes to the axial charge/current - lowest Landau level (LLL), while other LLs come in pairs canceling each other. Let us consider a flux tube with a strong constant magnetic field. Then the dynamics is separated for longitudinal and transverse directions. In 1+1d theory $E$ produces chiral modes running along $B$ and the transverse level density is defined by the magnetic flux

\begin{eqnarray}
&~&\partial_t N_{LLL}=\frac{1}{\pi}\int_C\vec Ed\vec x~~,~~\partial_t N_5=\frac{\int_C \vec Ed\vec x}{\pi}\frac{\int Bd^2x}{2\pi}\,.
\end{eqnarray}
Here $C$ corresponds to the contour along the (infinitely thin) tube and the last multiple in 3+1d relation is the transverse density of LLLs \cite{Landau:1991wop}. As it is mentioned above one  can still introduce a conserved axial charge despite the anomalous violation of the symmetry:

\begin{eqnarray}
\label{dq50}
\partial_t\left(N_5+\frac{1}{4\pi^2}\int A\cdot Bd^3x\right)=0\,.
\end{eqnarray}
This rephrasing of the axial anomaly reads: the change in the topological properties of the EM field configuration, magnetic helicity, is equal to the change in the number of zero modes on all flux tubes.

Recently this picture was used to obtain a realization of the CME current with no initial chiral asymmetry \cite{Hirono:2016jps}. It was shown there that in a reconnection process of two closed magnetic flux tubes an electric current of zero modes is generated along their contours $C_{1,2}$. Let us reproduce this calculation in somewhat reversed order. We first start with the integrated expression for the generalized axial charge (\ref{dq50}), which reads

\begin{eqnarray}
\Delta N_5+\frac{1}{4\pi^2}\Delta \mathcal{H}_{mh}=0\,,
\end{eqnarray}
assuming all processes to be localized in a finite volume. One notes that the generation of the axial charge in the strong $B$ limit corresponds to the production of zero modes. In 1+1d chiral theory two currents are related to each other $J_\mu^{1+1}=\epsilon_{\mu\nu}J_5^{1+1,\nu}$ and the presence of the axial charge unavoidably results in an electric current generated in the system. Indeed, zero modes are moving with speed of light along magnetic field and the direction of their momentum coincides with the $B$ direction. Thus, the full density of LLLs is

\begin{eqnarray}
&~&\Delta N_{LLL}=\oint_{\sum_iC_i} \Delta J_{1+1}\cdot dx
\end{eqnarray}
where integration is performed along all flux tubes in the system. Here $J_{1+1}$ is the $1+1d$  current density, and its absolute value is equivalent to the full electric current, that is charge transport per unit time through a cross section of a thin wire.
%  Noting that the 1+1d current density is just 3+1d current density integrated in the transverse plane one finds
%\begin{eqnarray}
%\label{hel}
%&~&\oint_{\sum_iC_i} \Delta I\cdot dx=-\frac{1}{4\pi^2}\Delta H_m
%\end{eqnarray}
%where $I$ is the electric charge flowing through a flux tube's cross section per unit time in the $B$ direction. 
It is worth mentioning that this derivation is mostly based on the conservation of the generalized axial charge and the result doesn't depend on details of the helicity changing process (cf. \cite{Hirono:2016jps}). For instance one may imagine two distinct ways to unlink two closed tubes possessing  no initial zero modes: ``shutting down'' one of flux tubes and then ``turning" it ``on" far away from the first one vs. a symmetric unlinking process. In the former case all zero modes are produced on the static tube due to the electric field generated by the change in the flux through the encircled area as per Faraday's law. In other words, the zero modes are generated due to the other tube ``shutting down" (the change in the flux). In the latter case zero modes have to be distributed symmetrically due to the geometry of the setup.

Magnetic field helicity corresponding to the linkage of closed flux tubes may be expressed as

\begin{eqnarray}
\mathcal{H}_{mh}=\sum_i S_i\Phi_i^2+2\sum_{i,j}L_{ij}\Phi_i\Phi_j,
\end{eqnarray}
where $\Phi_i$ is the magnetic flux of the $i$th tube, $L_{ij}$ is the Gauss linking number and $S_i$ is the Calugareanu-White self-linking number \cite{Moffatt, Moffatt:1992}. For a single reconnection one finds $\Delta \mathcal{H}_{mh}=2\Phi_1\Phi_2$ and we readily obtain

\begin{eqnarray}
\label{nCME}
\oint_{\sum_iC_i} \Delta J_{1+1}\cdot dx=-\frac{1}{2\pi^2}\Phi_1\Phi_2
\end{eqnarray}
in full agreement with \cite{Hirono:2016jps}\footnote{Note that in the formula (13) of \cite{Hirono:2016jps} %there is a wrong coefficient connected with somewhat sloppy notations and the resulting answer contains an incorrect factor of two
one should understand $2\phi_1\phi_2$ as $\phi_1\Delta\phi_2+\phi_2\Delta\phi_1$ which in turn is equal to $\frac{1}{2}\Delta H_{mh}$. As we show here this more general form is explicitly correct. However one may note that initially there is a factor of two difference caused by notations. This disagreement could be resolved via the axial charge conservation which leads to the result (\ref{nCME}) of our text.}.

\section{Chiral vortical effect}

At the present moment there is no microscopic description of the fluid velocity in terms of the underlying fields. Indeed, that would require dealing with highly nonlinear and nonperturbative processes. Moreover zero mode realization of chiral effects is under question in the hydrodynamic limit -- there are no classical fermionic fields and the anomaly must be realized by collective excitations of the medium (see, e.g., \cite{Sadofyev:2010is}). These complications prevent us from studying the microscopic details of the CVE phenomenon in a normal fluid. However there is a textbook example of a hydrodynamic system well described microscopically -- superfluid at zero temperature. In this section we adopt this idea to study the microscopic realization of CVE. An example of a chiral superfluid, for instance, is given by the CFL phase of baryon rich matter where some of the chiral effects were considered in \cite{Son:2004tq}. Here we adopt a rather general formalism referring interested reader to \cite{Son:2004tq} and references therein for particular models.

In a superfluid the velocity field is potential and related to the transformation parameter of the spontaneously broken symmetry group -- Goldstone boson \cite{landau1959fluid}. Thus one may write the Hamiltonian describing this system in terms of the microscopic superfluid potential avoiding the issue mentioned above. At the first glance the vortical effect is expected to be forbidden since the irrotational nature of the potential flow. However if one starts rotating some volume with a superfluid, at angular velocity higher than the critical one, it is energetically preferable to produce a defect carrying angular momentum \cite{landau1959fluid}. This superfluid vortex results in a singular behavior of the Goldstone field, particularly $\nabla\times v_s\neq0$. Thus, both angular velocity and CVE are localized on the defects \cite{Kirilin:2012mw, Son:2004tq}. The vortical contribution to the axial current in this setup is considered in \cite{Kirilin:2012mw}. It is shown that the current is a manifestation of the zero modes flow along a vortex\footnote{It should be mentioned that zero modes travel along the defect with the speed of light and are not thermalized with the medium. As shown in \cite{Kirilin:2012mw}, that results in the different value of CVE axial conductivity which is twice of its usual value.}.This picture coincides with the realization of the chiral separation effect in a magnetic flux tube \cite{Metlitski:2005pr}. Note that in general the anomalous dynamics is known to be closely related to zero modes on defects \cite{Nielsen:1983rb, Goldstone:1981kk, Callan:1984sa}. 

In hydrodynamic approximation a relativistic superfluid at $T=0$ is described by the corresponding scalar potential $\phi$ (see, e.g., \cite{Son:2002zn, Nicolis:2011cs} and references therein) which defines the chemical potential $\partial_t\phi=\mu$ and the fluid velocity $v_i^s=\partial_i\phi/\mu$ (or $u^s_\mu=\partial_\mu\phi/|\partial\phi|$). As discussed above, to introduce vortical effects into the system one first has to construct a superfluid defect. The simplest vortex solution, in the form of a straight line along the z-axis is given by $\phi=\mu t+\varphi$ where $\varphi$ is the polar angle in the transverse (xy)-plane \cite{Nicolis:2011cs, Landau:1980mil2}. One can directly check that $[\partial_x,\partial_y]\phi=2\pi\delta(x,y)$ and consequently $(\nabla\times \mu v_s)_z=2\pi\delta(x,y)$. It should be noted that disregarding the vortex configuration the angular momentum is conserved resulting in $\int \vec\partial\phi d\vec x=2\pi n$, where $n$ is the winding number for the given defect.

The fermionic zero modes on a defect could be described by a Dirac action in the presence of the Goldstone field of the spontaneously broken symmetry (see \cite{Kirilin:2012mw}):

\begin{eqnarray}
\label{action}
S_f=\int d^4x~i\bar\psi\left(\partial_\mu+i\partial_\mu\phi\right)\gamma^\mu\psi\,.
\end{eqnarray}
This picture simplifies the consideration since one can treat medium effects on the fermionic modes just in terms of a slowly varying external field.

One expects that the superfluid potential $\phi$ enters the action in the same way as a pure gauge for the electromagnetic field. Then, the angular momentum possessed by the vortex plays the role of magnetic flux, while electric field should be replaced by perturbations in the chemical potential. Thus, we anticipate that

\begin{eqnarray}
\label{sfq5}
\partial_t\left(N_5+\frac{1}{4\pi^2}\mathcal{H}_{sfh}\right)=0\Rightarrow \oint_{\sum_i C_i} \Delta J_{1+1}\cdot dx=-\frac{1}{4\pi^2}\Delta \mathcal{H}_{sfh}
\end{eqnarray}
where $\mathcal{H}_{sfh}$ is the superfluid flow helicity given by a substitution of $\vec A\to \mu\vec{v}_s$ to the magnetic helicity.

If there is a linkage between two superfluid vortices $C_{1,2}$ (with $n_1$ and $n_2$) then $\mathcal{H}_{sfh}\neq0$ and it can be expressed in terms of the linking number similarly to the magnetic case. Note that it is expected since two conservation laws coincide at the algebraic level and the only difference appears in the interpretation of the field entering the corresponding helicity ($\vec A\leftrightarrow \mu \vec{v}_s$). Indeed, one may write for the commutator of two derivatives acting on $\phi$:

\begin{eqnarray}
\label{sfB}
\epsilon^{abc}\partial_b\partial_c\phi=2\pi n_1\int\delta^{(3)}(x-z_1(s))\frac{dz^a_1(s)}{ds}ds+2\pi n_2\int\delta^{(3)}(x-z_2(s))\frac{dz^a_2(s)}{ds}ds\,.
\end{eqnarray}
Here $\Gamma_i$ is a contour encircling $C_i$ and we use $\int_{\Gamma_i} \vec\partial\phi d\vec x=2\pi n_i$. The $i$th defect is given by $z_i(s)$ with $s$ being the length parameter. Note that the angular momentum possessed by a vortex is in one to one correspondence with the winding number. The expression above may be understood as the angular momentum conservation along a vortex line. Substituting (\ref{sfB}) into the magnetic helicity one finds

\begin{eqnarray}
\label{2vH}
\mathcal{H}_{sfh}=\int d^3x \mu v_s\cdot(\nabla\times \mu v_s)=2\pi n_1\oint_{C_1}\partial \phi\cdot dx+2\pi n_2\oint_{C_2}\partial \phi\cdot dx=8\pi^2n_1n_2\,.
\end{eqnarray}
Note that this result can be easily obtained from a simple substitution $\Phi_i\to 2\pi n_i$ into $\mathcal{H}_{mh}$, resulting in $\mathcal{H}_{sfh}=8\pi^2n_1n_2$. Thus we expect that if an unlinking of two lines happens it results in a zero mode current along their contours satisfying

\begin{eqnarray}
\label{nCVE}
 \int_{C_1+C_2} \Delta J_{1+1}\cdot dx=-2n_1n_2\,.
\end{eqnarray}
This quantized CVE current is analogous to (\ref{nCME}) and could be readily generalized to the case of multiple vortices. Note that the distribution of the current between contours depends on details of the process.

It is worth mentioning that the fundamental theory has no anomaly in the absence of external fields. The source of the anomaly here is the superfluid potential $\phi$ entering as a pure gauge into the Lagrangian (\ref{action}). The presence of the defects makes this field multivalued and one expects it to generate a nonzero ``magnetic'' field in the vortex. Then, the ``electric field'' sourced by the chemical potential gradient saturates the anomalous divergence. One can argue that in the present formalism there is no visible difference between the electric field and its flow analogue. However one should remember that the chemical potential can be of a nonelectromagnetic nature. %, e.g., for the pionic superfluid where chemical potential corresponds to the global charge - isospin.

This picture can be illustrated microscopically through the unlinking process of two superfluid vortices. Let us suppose that one of defects is a circle (with the winding number $n_1$) while another one is a straight line perpendicular to the circle and crossing it at the center. Then, according to (\ref{2vH}), the full flow helicity of the system is

\begin{eqnarray}
\mathcal{H}_{sfh}=8\pi^2n_1n_2\,.
\end{eqnarray}
If we change the angular momentum possessed by the straight vortex, or equivalently change $n_2$, it results in the flow change along the closed vortex

\begin{eqnarray}
2\pi\frac{\Delta n_2}{\Delta t}=\frac{\Delta\left(\oint_{C_2} \vec\partial\phi d\vec x\right)}{\Delta t}\,,
\end{eqnarray}
where $\Delta$ is used to stress that the change in $n_2$ must be quantized. The jump of the circulation unavoidably results in a jump of the superfluid potential $\phi$ which could be considered as a kink-like time dependence. Smoothing out this jump we conclude that zero modes are created along the closed vortex according to the 1+1d anomaly of the given Lagrangian. The transverse density in the closed loop stays constant and it is given by $n_1$ (compare with the magnetic case). Combining two pieces together we find

\begin{eqnarray}
\frac{d}{dt} N_{5}=-\frac{n_1}{\pi}\frac{d}{d t}\left(\oint_c \vec\partial\phi d\vec x\right)=-2n_1\frac{dn_2}{dt}
\end{eqnarray}
and, since $n_1=const$, one may see that $\frac{d}{dt} N_{5}=-\frac{d}{dt}(2n_2n_1)$. Recalling that $\mathcal{H}_{sfh}=8\pi^2n_1n_2$ we finally obtain the vortical analogue of (\ref{nCME}) which is given by

\begin{eqnarray}
\label{nCVEanother}
 \int_{C_1+C_2} \Delta J_{1+1}\cdot dx=-\frac{1}{4\pi^2}\Delta \mathcal{H}_{sfh}\,,
\end{eqnarray}
as expected on the basis of the conservation law (\ref{sfq5}). This simple exercise provides a direct check of the zero modes generation by a change in the macroscopic helicity of the flow. The relation between superfluid velocity and the field of the broken group explains the novel anomalous contribution which can be attributed to the anomaly constructed from the ``initial" unbroken fields. Following the terminology of \cite{Hirono:2016jps} one can refer to this effect as to the quantized CVE or equivalently say that we generate chiral asymmetry by a change in the superfluid flow helicity. It should be mentioned that a reconnection requires the presence of some dissipative process. However, even in a superfluid, vortices do interact with each other and the normal component, allowing the helicity to change.

It is instructive to study a similar situation with the usual electric field $E$. Then 1+1d anomaly responsible for the zero modes production is due to $E$ and not $\partial_t\mu\vec{v}_s$. Therefore if there is a change in the linkage of a superfluid vortex and a magnetic flux tube it results in the zero mode production, in general, on both defects. Such process takes place as long as the modes are charged with respect to both fields. Let us consider the cross magnetic-flow helicity of a linkage between a superfluid vortex $C_s$ and a flux tube $C_t$, ignoring its specific realization in the medium:

\begin{eqnarray}
&~&\mathcal{H}_{mfh}=2\pi n\int_{C_s}A(z_s)\cdot dx+\Phi\int_{C_t}\nabla\phi\cdot dx=4\pi n\Phi\,.
\end{eqnarray}
Recalling Fraday's law and its analogue for the superfluid flow we conclude, as previously, that the reconnection results in

\begin{eqnarray}
\label{mixednCE}
 \int_{C_t+C_s} \Delta J_{1+1}\cdot dx=-\frac{1}{\pi}n\Phi\,.
\end{eqnarray}
Thus one may generate both CME or CVE currents through a change in the cross helicity $\mathcal{H}_{mfh}$. This result combined with the generation of zero modes through a change in the other helicities completes the full set of chiral effects caused by reconnections. It also provides a direct probe of the axial charge unification \cite{Avdoshkin:2014gpa} in the case of chiral superfluid.

\section{Competing processes}

In this section we concentrate on the hydrodynamic limit of magnetic reconnections in chiral medium.  One may note that in the exact chiral limit there is no mechanism to dissipate the current of zero modes flowing along flux tubes, it is persistent \cite{Kharzeev:2011ds, Sadofyev:2015tmb}. Indeed, if one starts with a state of a nonzero helicity the produced number of zero modes is proportional to the change in the helicity from the initial until the current moment. Moreover since finite conductivity results in continuous dissipation of the helicity through reconnections the final current is seemingly constrained only by the initial $\mathcal{H}_{mh}$ value and its spatial distribution dynamics. %\comm{\bf and dynamics of the re-distribution in space}.

This unstable behavior, with the system falling into the specific state with all the helicity transferred into the fermionic chiral asymmetry is challenged. There is a competing process turning the asymmetry into helical magnetic field. Indeed, if one starts with zero magnetic helicity and non-zero chiral asymmetry, chiral medium is known to be unstable \cite{Boyarsky:2012ex, Akamatsu:2013pjd}. Moreover such instability is a general feature of systems with electric current directed along magnetic field $\vec J=\alpha\vec B$. For instance consider the $\alpha$-dynamo in turbulent magnetohydrodynamics (MHD) \cite{Moffatt}, where the instability is sourced by the large-scale electric current which, in turn, is caused by the small-scale P-odd turbulence (for recent review see \cite{Yokoi:2013mja}). 

Let us discuss the simplest possible model of the instability ignoring $(\omega, k)$-dependence of the conductivity and $\alpha$. We begin with introducing the helical decomposition of the magnetic field, with respect to the eigenmodes of the curl operator (the CK states):
\be
\nabla\times \vec{B}=K\vec{B}
\ee
where $K$ is the corresponding eigenvalue. Then the MHD equation takes the following form:
\begin{eqnarray}
&~&\frac{\partial}{\partial t}\vec B=\frac{1}{\sigma}\Delta \vec B+\frac{\alpha}{\sigma}\nabla\times \vec B\Rightarrow\notag\\
&~&\frac{\partial}{\partial t}\vec B=-\frac{K^2}{\sigma} \vec B+\frac{\alpha K}{\sigma}\vec B
\end{eqnarray}
and one can readily see that modes with $K<\alpha$ are growing in time. These solutions are thought to describe stellar and galactic magnetic fields (for review see \cite{Giovannini:2003yn, Yokoi:2013mja}). Note that in the limit $\sigma\to\infty$ both the magnetic helicity and the chiral asymmetry are conserved due to complete screening of the electric field, in agreement with the equations above. It is usually argued that the final state of the system driven by this instability should be determined by the non-linearities which become relevant as the magnitude of the magnetic field increases. In our case the conductivity coefficient $\alpha$ is expected to be $B$-dependent, as it is determined by the helicity according to \eqref{nCME}. The final state corresponds to the requirement 

\begin{eqnarray}
\vec E=0~\Rightarrow~\nabla\times \vec B-\alpha \vec B=0\,.
\end{eqnarray}
%f the initial state has instead zero magnetic helicity and non-zero chiral asymmetry, then the system tends to generate helicity lowering $N_5$. %On the other hand, for a helical initial state with no chiral imbalance the helicity dissipates into the asymmetry due to the reconnection processes. 

To reiterate, let us address this picture of two competing processes in chiral plasmas in more details relying on energy balance and helicity conservation(see \cite{Hirono:2015rla, Zakharov:2016lhp}). The charge $N_5$ is not conserved while there is a constraint (\ref{dq50}). In the state with the entire axial charge represented by $N_5$ the energy could be lowered by creating arbitrarily soft CK modes, and thus transferring of the chiral asymmetry to the magnetic helicity\footnote{Note that the magnetic energy is not determined with the helicity but rather bounded by its value, one expects the system to follow the minimal energy configuration possessing the required helicity.}.  In the opposite setup of no initial $N_5$, the chemical potential is zero since on general grounds $\mu_5 \propto N_5$ for a system close to equilibrium, and one can lower energy decreasing $B$ and generating $N_5$. The final helical field configuration should be supported by CME current which requires some finite chiral imbalance\footnote{This is due to the fact that the vacuum Maxwell equations do not support a static solution with a finite helicity. For an explicitly time-dependent solution, see \cite{Ranada:1989wc}.}.

Studying the specific endpoint of the instability requires detailed knowledge on the dynamics of two processes. In \cite{Hirono:2015rla} it is shown that for an anomalous MHD model this state is fixed by the system size, which determines the lowest eigenvalue $K_{min} \sim 1/L$ of the CK modes. Indeed, as mentioned above, the energy per helicity could be lowered by transfer of the microscopic asymmetry to the softest possible CK mode. This solution is self-consistent, the residual chiral imbalance supports this field configuration and in larger and larger system it tends to zero. On the other hand it is argued in \cite{Zakharov:2016lhp} that in the absence of large intrinsic parameter the axial charge should be somewhat equally distributed among available forms.
%In other words the ratio of microscopic contribution to the axial charge to the macroscopic one is zero in the thermodynamic limit. %

To resolve this seeming disagreement we note that the picture above indicates the final distribution to be IR-sensitive. This is a general feature of the anomalous transport (see, e.g., \cite{Kirilin:2013fqa, Khaidukov:2013sja}) and the relative magnitude of IR parameters smallness should be specified. It is worth mentioning that the anomaly can be treated as an IR phenomenon \cite{Dolgov:1971ri} which is saturated on the relevant scale. Thus it is reasonable to expect dependence of the instability end point on other IR parameters, say the small mass of the medium constituents. Then, the picture given in \cite{Hirono:2015rla} may be considerably modified and it is compatible with arguments \cite{Zakharov:2016lhp} since the applicability of the chiral limit contains assumptions on the smallness of the mass (even at the hydrodynamic scale) \cite{Kirilin:2013fqa}. If the time scale of chirality dissipation due to finite mass of the fermions is shorter than the relevant scale of helicity-to-chirality transfer the softest configuration has no required supporting current along magnetic field. The system tends to a stabilized regime with the minimal energy and it is natural that it would be determined by the balance of two processes (chirality decay and helicity-to-chirality transfer). For the field configuration with smaller characteristic length scale of magnetic field nonuniformity, the rate of $\mathcal{H}_{mh}\to N_5$ is larger. %\com{and depends also on the electric conductivity $\sigma$ (the Ohm's dissipative part)}. 
Thus the final distribution of the full axial charge in a large enough system would be rather fixed by the mass than by the system size along this oversimplified line of arguments. We conclude that the endpoint of the instability depends on the full set of IR parameters.%However quantitative analysis is required and we leave that for future study.

Similarly, one can combine recently predicted vortical instability \cite{Avdoshkin:2014gpa, Zakharov:2016lhp} with the CVE caused by reconnections of superfluid vortices. Then, these two competing processes redistribute the generalized axial charge among available degrees of freedom. However more detailed analysis of the vortical instability requires to deal with highly nonlinear dynamics of defect production \cite{Burch:2015mea}.

\section{Discussions}

In this paper we study the generalization of the axial charge by EM fields and chiral medium motion. The axial symmetry present at the classical level is known to be violated by loop contributions \cite{Adler:1969gk, Bell:1969ts}. However one may introduce a redefined axial charge (\ref{q50}) conserved in external fields (\ref{dq50}). This charge is shifted by the macroscopic helicity of magnetic field and the conservation may be extended to the dynamical situation. Recently it was proposed  \cite{Avdoshkin:2014gpa} that in a medium there are additional vortical counterparts in (\ref{q50}). All macroscopic contributions to the generalized axial charge are of similar topological nature and correspond to the linking numbers between field and flow lines (\ref{q5}).

However the microscopic origin of the generalized conservation law (\ref{q5}) is puzzling. Indeed, the vortical contributions to (\ref{CE}) are present in the system even in the absence of EM fields. Thus, there is a modification of the axial charge by the flow helicity while there is no anomaly at the microscopic level. Indeed, the axial anomaly gains no corrections in the medium and disappears along with EM fields. Further studying of the microscopic details of the vortical contributions is additionally complicated by the highly nonperturbative nature of the classical notion of ``velocity'' in relation to the microscopic fields. Here we concentrate on a toy model of chiral supefluidity which provides a microscopic picture for the macroscopic medium motion \cite{Son:2002zn, Nicolis:2011cs}. 

At zero temperature superfluid is known to be irrotational since the flow is potential. On the other hand after some critical angular velocity a rotating volume with a superfluid transfers some angular momentum to the medium \cite{Landau:1980mil2}. That results in a change of the vacuum which, in order to carry some angular momentum, has to contain defects: superfluid vortices. Thus the CVE can be realized in the system but has to be localized along a vortex. Linking of several vortices generates macroscopic helicity $\mathcal{H}_{sfh}$ which is expected to contribute into the generalized axial charge (\ref{q5}). We study this setup and explicitly check that the superflow helicity can be transferred to the fermionic zero modes. It provides a specific microscopic realization of the vortical contributions to the axial charge. It should be mentioned that this modification is expected since the superfluid potential cannot be separated from the Goldstone field which in turn can contribute to the anomaly. Turning to the multiple vortex limit one may simulate rotation of a normal fluid. However this phenomenological transition between two pictures requires further study.

We apply the resulting conservation law for the generalized axial charge (\ref{q5}) to derive the result for the anomalous transport caused by the transfer of the superfluid flow helicity to the fermionic chirality (\ref{nCVEanother}). The novel contribution to the transport is a quantized CVE current. The analogous effect in the case of magnetic reconnections was recently studied in \cite{Hirono:2016jps}. This allows us to predict similar effects caused by a change in the mixed helicity counting the linkage of the flow lines with magnetic field lines (\ref{mixednCE}). Thus in our model there is a wider set of chiral effects caused by the changes in the helical part of the axial charge and requiring no initial chiral asymmetry.

The averaged macroscopic currents caused by helicity transitions modify the dynamics of chiral plasma. These effects by their nature are similar to dynamo phenomena (see, e.g., \cite{Moffatt}) known to be important in MHD turbulence. We consider the competition between the transfer of the magnetic helicity to the chiral asymmetry and the chiral instability (see, e.g., \cite{Boyarsky:2012ex, Akamatsu:2013pjd}). In \cite{Avdoshkin:2014gpa, Zakharov:2016lhp} the axial charge is argued to be somewhat equally distributed among different forms. On the other hand in \cite{Hirono:2015rla} it is shown that the final distribution is asymmetric since one may lower energy at given axial charge by transferring the microscopic asymmetry to the soft helical modes of the magnetic field (specifically the softest CK mode). Such transfer is constrained only by the size of the system and the residual chiral asymmetry, required to support the field configuration, tends to zero. Resolving this disagreement we argue that in the consideration of \cite{Hirono:2015rla} there is an internal small parameter. Indeed, considering the chiral limit one has to suppose the constituent mass to be small even at the hydrodynamic scale. Chiral effects are known to be IR sensitive \cite{Kirilin:2013fqa, Khaidukov:2013sja} and it is natural to expect that the final distribution depends on the complete set of IR parameters. For instance, in the limit of a large system $L\to\infty$ at some point the chirality flipping process becomes important when $m\sim1/L$. The softest mode cannot be supported anymore and it is natural to assume that the system will stabilize in the intermediate state when two competing  time scales are close. Thus we conclude that in the general setup the endpoint of the chiral magnetic instability depends on the full set of IR parameters. A quantitative study of the final distribution requires detailed knowledge of the IR dynamics in a specific system and presents a challenge for the future numerical simulation.

Finally we stress that currents caused by the axial charge conservation provide a competing process to the general set of chiral instabilities \cite{Avdoshkin:2014gpa}, including the chiral vortical instability \cite{Zakharov:2016lhp}. However in the latter case the analysis is complicated by nonlinear realization of this process through vortex production and we leave it for the future work. It should also be mentioned that in the case of the chiral magnetic instability the final state is modified by the mixed reconnections and we propose to study this process in a full MHD consideration.

\section{Acknowledgments}
We  are  grateful  to D.  Kharzeev,  K. Rajagopal and Y. Yin for useful discussions.  We especially want to thank V.I. Zakharov  for illuminating remarks and the motivation for this work.  The work of AS was supported in part by the U.S. Department of Energy under grant Contract Number DE-SC0011090. The work of VK was supported
in part by the US NSF under Grant No. PHY-1620542 and in part by RFBR grant 17-02-01108А. The work on the sections III and IV is supported by Russian Science Foundation Grant No. 16-12-10059.

\bibliographystyle{ssg}
\bibliography{limit_v.bib}

\end{document}